\documentclass[aip,apl,reprint]{revtex4-1}
\usepackage{epsf,graphicx}% Include figure files
\usepackage{amsmath,amsfonts,amssymb}
\usepackage{subeqnarray}% Number equations with letter
\usepackage{dcolumn}% Align table columns on decimal point
\usepackage{bm}% bold math
\usepackage{xcolor}
\usepackage{ulem}
\usepackage{mathrsfs}
\usepackage{multirow}
\usepackage{array}
\newcommand{\ra}[1]{\renewcommand{\arraystretch}{#1}}
\newcommand{\head}[1]{\multicolumn{1}{>{\centering\arraybackslash}p{#1}}}
\begin{document}
\title{A generic tight-binding model for monolayer, bilayer and bulk MoS$_{2}$}
\bigskip

\author{Ferdows Zahid}
\email{fzahid@hku.hk}
\affiliation{Department of Physics, The University of Hong Kong, Pok Fulam Road, Hong Kong SAR, China}

\author{Lei Liu}
\affiliation{Nanoacademic Technologies Inc., Brossard, Quebec, J4Z 1A7, Canada}

\author{Yu Zhu}
\affiliation{Nanoacademic Technologies Inc., Brossard, Quebec, J4Z 1A7, Canada}

\author{Jian Wang}
\affiliation{Department of Physics, The University of Hong Kong, Pok Fulam Road, Hong Kong SAR, China}
\author{Hong Guo}
\affiliation{Centre for the Physics of Materials and Department of Physics,
McGill University, Montreal, PQ, H3A 2T8, Canada}

\date{\today}% It is always \today, today,

\medskip
\widetext
\begin{abstract}
Molybdenum disulfide (MoS$_{2}$) is a layered semiconductor which has become very important recently as an emerging electronic device material. Being an intrinsic semiconductor the two-dimensional MoS$_{2}$ has major advantages as the channel material in field-effect transistors. In this work we determine the electronic structure of MoS$_{2}$ with the highly accurate screened hybrid functional within the density functional theory (DFT) including the spin-orbit coupling. Using the DFT electronic structures as target, we have developed a single generic tight-binding (TB) model that accurately produces the electronic structures for three different forms of MoS$_{2}$ - bulk, bilayer and monolayer. Our TB model is based on the Slater-Koster method with non-orthogonal sp$^{3}$d$^5$ orbitals, nearest-neighbor interactions and spin-orbit coupling. The TB model is useful for atomistic modeling of quantum transport in MoS$_{2}$ based electronic devices.
\end{abstract}
\bigskip
\maketitle

Molybdenum disulfide (MoS$_{2}$) belongs to a family of layered transition metal dichalcogenides (TMDC) in which the layers are held together by weak van der Waals forces, and it can be exfoliated mechanically to a single layer thickness. In its bulk form MoS$_{2}$ is an indirect band gap semiconductor which turns into a direct band gap semiconductor for monolayer structure.\cite{rHeinz} This intrinsic semiconducting nature of MoS$_{2}$ is a major advantage over graphene (which has no intrinsic band gap) as a two-dimensional (2D) channel material in field-effect transistors (FET). Indeed, FET devices based on MoS$_{2}$ monolayer and bilayer have already been fabricated in the experimental labs and demonstrated to have useful device performances.\cite{rKis, rLiu} Several theoretical studies of MoS$_{2}$ FET devices \cite{rSayeef, rGuo, rKalam} based on simplified description of the electronic structures within an effective mass model have also been reported recently.

Due to the complex nature of the layered TMDC materials and the great interests in the electronics community for its applications in emerging devices,  a more reliable, accurate, and atomistic treatment of the electronic structures of TMDC is desired. Since \textit{ab initio} models are computationally expensive and often intractable for realistic device structures having large number of atoms, a widely applied intermediary option is the tight-binding (TB) model. TB models are atomistic full-band yet computationally efficient due to the small number of basis orbitals and the rigid nature of the Hamiltonian. Properly parameterized TB models have been extensively applied to predict reliable results for a diverse range of devices including the resonant tunneling diodes, \cite{rRoger} quantum dots, \cite{rGerhard} Si nanowireFET, \cite{rMathieu} and compound semiconductor heterostructures.\cite{rIntel} However, for the very exciting 2D MoS$_{2}$ and other TMDC materials, accurate TB models for electronic structures has not yet been developed. It is the purpose of this work to fill this void, and the resulting TB model should be extremely useful for predicting quantum transport properties of MoS$_{2}$ based devices.

In particular, we have developed a \textit{generic} TB model with non-orthogonal sp$^{3}$d$^5$ orbitals including the nearest-neighbour interactions and spin-orbit coupling, that accurately determines the electronic structures of bulk, monolayer and bilayer MoS$_{2}$. A reliable and accurate target band structure is the primary requirement for a successful tight-binding modeling. In this study, for the calculations of the target band structures we employ density functional theory (DFT) with the screened hybrid functional of Heyd, Scuseria, and Ernzerhof \cite{rHeyd1} (HSE) that has been shown in the literature \cite{rKresse1, rKresse2, rHyd2} to produce accurate band gaps and reasonable effective masses for a wide range of semiconductors. Our calculated band structures of MoS$_{2}$ from the self-consistent DFT-HSE shows excellent agreement with the available experimental data and compare well with other theoretical studies.\cite{rLamb, rEllis, rWalle} Even though an accurate target band structure is obtained by DFT-HSE, it is nontrivial to develop a \textit{generic} TB model which can accurately capture electronic structures and effective masses of not only bulk MoS$_{2}$ but also its double-layer and single-layer forms. For the layered TMDC structures there is a thickness dependent interlayer interaction that significantly affects the electronic properties, the TB model must capture these effects so as to be applicable to devices made of different layer thicknesses. As shown below, our \textit{generic} TB model accurately produces the band structures for three different forms of MoS$_{2}$: bulk, monolayer and bilayer. The reported TB model can be easily extended to other MoS$_{2}$ structures of higher number of layers  as well as other layered TMDC materials such as WS$_2$, WSe$_2$, and MoSe$_2$. The accuracy of our TB model is validated by comparing the band gaps and effective masses with our calculated \textit{ab initio} results.

\begin{center}
%\vspace{-0.3cm}
\begin{figure}[h]
{\epsfxsize 8.0cm \epsfbox{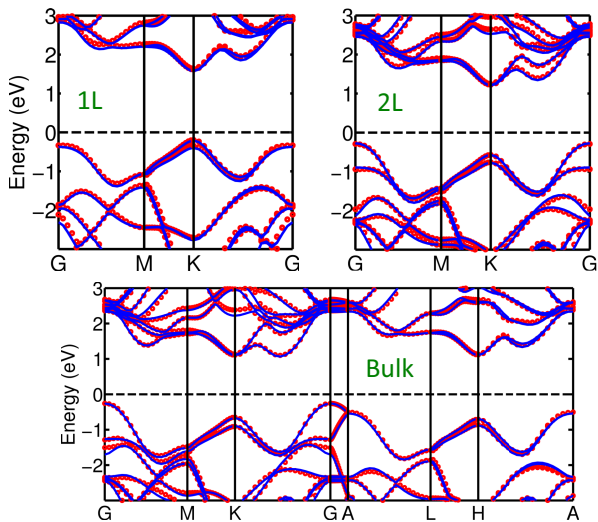}} \vspace{0.1cm}
\caption{(color online) Band structures for monolayer ($1$L), bilayer ($2$L), and bulk MoS$_{2}$ using screened hybrid density functional theory (blue lines) and our tight-binding fitting (red dots). The zero in the energy axis is set at the Fermi level as shown by the dashed line. Spin-orbit coupling is included in the calculations
self-consistently.}
\label{fig1}
\end{figure}
\end{center}

In our study, all the DFT-HSE calculations are performed using the projector augmented-wave (PAW) pseudopotential plane-wave method as implemented in the VASP software package.\cite{rVasp} A Monkhorst-Pack scheme is adopted for the sampling of the Brillouin zone with k-point meshes of $7 \times 7 \times 2$ for the bulk and $9 \times 9 \times 1$ for the two-dimensional (2D) structures. An energy cutoff of $280$ eV is used in a plane wave basis set. We also included the spin-orbit coupling self-consistently. Due to the presence of van der Waals interaction which determines the interlayer distance in MoS$_{2}$, geometry optimization becomes quite tricky and complicated. Without the van der Waals intereaction the lattice parameter $c$ (vertical length of the unit cell) is overestimated by a large margin. However, it is not yet possible to include van der Waals interaction in the HSE calculations, and it is an area of active research. We therefore optimize the bulk crystal structure using the  Perdew-Burke-Ernzerhof (PBE) functional \cite{rPBE} for which the van der Waals interaction can be included explicitly by adopting the DFT-D2 method. \cite{rvdW} We then use this optimized structure for the HSE calculations. The values of our optimized lattice parameters for bulk  MoS$_{2}$ are: $a = 3.179$ {\AA} and $c = 12.729$ {\AA}  with a layer thickness of $3.135$ {\AA} (S-S vertical distance). Although the value of $c$ parameter improves significantly with the DFT-D2 method, it is still overestimated by around $3.6\%$ compared to the experimental value.

\begin{center}
\begin{figure}[h]
{\epsfxsize 8.0cm \epsfbox{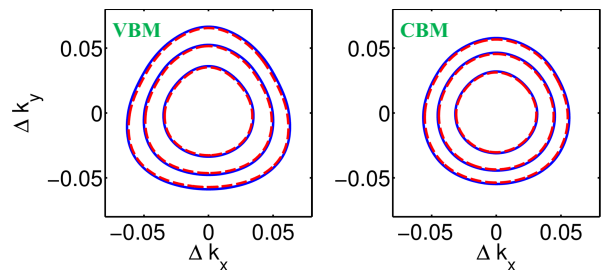}} \vspace{0.1cm}
\caption{(color online) Contour plots of the energies around the valence band maximum (VBM) and conduction band minimum (CBM) for monolayer MoS$_{2}$ from screened hybrid density functional theory (blue solid lines) and our tight-binding model (red dashed lines). The contour lines are associated with $\Delta E = \pm \frac{1}{30}, \pm \frac{1}{15}, \pm \frac{1}{10}$ eV, respectively and the $\Delta k_x$ and the $\Delta k_y$ are in the unit of $\frac{2\pi}{a}$, where $a$ is the unit cell length.}
\label{fig2}
\end{figure}
\end{center}

It is a well-known fact that DFT within the local density approximation (LDA) or the generalized gradient approximation (GGA) underestimates band gaps of semiconductors due to factors such as the self interactions. This error can be minimized in the HSE scheme of hybrid functional method\cite{rHeyd1} by incorporating $25\%$ short-range (SR) exact Hartree-Fock (HF) exchange with the PBE functional. A screening parameter $\mu$ defines the range separation, and is empirically set to $0.2$ {\AA}$^{-1}$ (HSE06 scheme) for both the HF and PBE parts. In our calculations for MoS$_{2}$ we observe that the band gap energies vary significantly with the screening parameter for both the bulk and the $2$D structures. We therefore decided to adjust the value of $\mu$ to fit the experimental band gap, and an optimize value for $\mu$ is found to be $0.4$ {\AA}$^{-1}$ which corresponds to an interaction range of $\pi/\mu = 7.85$ {\AA} for the SR nonlocal exchange. Note that a single value of $\mu$ is used for all the three structures of MoS$_{2}$: bulk, double-layer and single-layer, namely the HSE results presented here were all obtained using $\mu = 0.4$ {\AA}$^{-1}$.

Our calculated band structure results for MoS$_{2}$ are plotted in Fig.~\ref{fig1}. We notice that the monolayer MoS$_{2}$ has a direct band gap at the K point. However, with the addition of just one extra layer (i.e. for a bilayer structure) an indirect band gap opens up due to the presence of interlayer interactions. Although both bulk and bilayer MoS$_{2}$ have indirect band gaps the conduction band minima (CBM) are located at different points: for bulk it is at $\Sigma$ point (midpoint of $\Gamma$ and K points) and for bilayer it is at K point. Similar behavior is observed in a previous theoretical study. \cite{rLamb} For monolayer MoS$_{2}$ the splitting of the valence band maximum (VBM) at K point is solely due to the spin-orbit coupling (SOC) whereas a combination of SOC and interlayer interactions is responsible for the VBM splitting in bulk and bilayer MoS$_{2}$. The splitting of CBM due to SOC is minimal (around $5$ meV). The calculated band gap energies for different transitions are presented in Table~\ref{table1}. Our DFT-HSE results show good agreement with the experimental data. The values of effective masses at different band edges along different directions as presented in Table~\ref{table2} also compare well with other available theoretical studies. \cite{rLamb, rWalle}

\begin{table}[h]
\caption{Band gap energies obtained by DFT-HSE and our TB model. The fifth column is the deviation between the HSE and the TB values. Experimental data is shown in the sixth column, taken from Ref. \onlinecite{rbulk} (for bulk) and Ref. \onlinecite{rHeinz} (for monolayer and bilayer). All the energies are in the unit of eV. Subscripts $v$ and $c$ stand for valence band and conduction band, respectively. The splitting of the valence band maximum at K point is given by K$_{v1}$ (top) and K$_{v2}$ (bottom), whereas $\Sigma$ is the midpoint of the line joining the $\Gamma$ and the K points.}
\ra{1.3}
 %\centering % used for centering table
\begin{tabular}{c c c c c c} % centered columns (6 columns)
\hline\hline %inserts double horizontal lines
Structure & Gap & \head{1.2cm} {HSE (target)} & \head{1.2cm}{TB (fitted)} & \head{1.5cm} {Deviation (\%)} & Exp. \\ [0.5ex] % inserts table heading
\hline % inserts single horizontal line
Monolayer & K$_{v1}$ to K$_{c}$ & 1.786 & 1.805 & 1.06 & 1.90 \\
               &  K$_{v2}$ to K$_{c}$ & 1.974 & 1.969 & 0.24 & 2.05 \\
Bilayer        &  $\Gamma_{v}$ to K$_{c}$ & 1.480 & 1.516 & 2.41 & 1.60 \\
               &  K$_{v1}$ to K$_{c}$ & 1.779 & 1.792 & 0.76 & 1.88 \\
              &  K$_{v2}$ to K$_{c}$ & 1.980 & 1.987 & 0.35 & 2.05 \\
Bulk          &  $\Gamma_{v}$ to $\Sigma_{c}$ & 1.328 & 1.331 & 0.22 & 1.29 \\
              &  K$_{v1}$ to K$_{c}$ & 1.776 & 1.749 & 1.46 & 1.88 \\
              &  K$_{v2}$ to K$_{c}$ & 1.960 & 2.009 & 2.46 & 2.06 \\ [0.1ex] % [1ex] adds vertical space
\hline %inserts single line
\end{tabular}
\label{table1} % is used to refer this table in the text
\end{table}

With the calculated DFT-HSE band structures as target, we employed the Nanoskif \cite{Nanoskif} software package to obtain a set of TB parameters for the on-site energies, the Slater-Koster energy integrals, overlap integrals, \cite{SK} and the spin-orbit splitting. \cite{Chadi} In our TB model we adopted a non-orthogonal basis set of sp$^{3}$d$^5$ orbitals considering only the nearest neighbour interactions. The effects of spin-orbit coupling is included through a split-off energy term. A set of 96 parameters are optimized through the automated process built in Nanoskif.\cite{Nanoskif} The root-mean-square deviation of the fitting is within
25 meV. The optimized TB parameters are listed in Table ~\ref{table3}. Note that this single set of TB parameters is capable of producing accurate band structures of all three different forms of MoS$_{2}$.

The band structures obtained from our TB model are plotted in Fig.~\ref{fig1} along with those obtained from DFT-HSE. The agreement in energies around both VBM and CBM for all three structures of MoS$_{2}$ is excellent. Fig.~\ref{fig2} shows the energy contours around the valence band maximum (VBM) and the conduction band minimum (CBM) for the monolayer MoS$_{2}$. Again, we observe excellent agreement in the shape of the energy contours from DFT-HSE and TB fitting. The band gap energies for different transitions and the effective mass values at different symmetry points obtained from the TB model are presented in Table~\ref{table1} and Table~\ref{table2} respectively. For the band gap energies a fitting accuracy of less than $2.5\%$ is achieved. On the other hand, for the effective masses, in most cases the deviations between the DFT-HSE and the TB values are less than $10\%$ which is quite acceptable. In four cases (excluding the monolayer structure) the deviations are higher with the worst situation at $\sim 27.9\%$. Given that just a single set of TB parameters is used to produce the band structurues of all three different forms of the material, this level of  quantitative consistency is rather satisfactory.

\begin{widetext}
\begin{center}
\begin{table}[h]
\caption{Values of effective masses at various band edges in the unit of free electron mass (m$_{0}$) calculated using the HSE method and our tight-binding model. The subscripts $l$ and $t$ refer to the masses calculated at the point along the longitudinal and the transversal directions of the line connecting the $\Gamma$ point and that point, respectively.}
\ra{1.3}
\begin{tabular}{c c c c c c c c} % centered columns (8 columns)
\hline\hline %inserts double horizontal lines
     &      & \multicolumn{3}{c}{Electron} & \multicolumn{3}{c}{Hole} \\
%\cline(l{.75em}r{.75em}){3-5}
%\cline(l{.75em}r{.75em}){6-8} \\
%\cline{3-5}
%\cline{6-8}
Structure & Point & \head{1.2cm} {HSE (target)} & \head{1.2cm}{TB (fitted)} & \head{1.5cm} {Deviation (\%)} &  \head{1.2cm} {HSE (target)} & \head{1.2cm}{TB (fitted)} & \head{1.5cm} {Deviation (\%)} \\ [0.5ex]
% inserts table heading
\hline % inserts single horizontal line
Monolayer & K$_{l}$ & 0.407 & 0.430 & 4.58 & 0.485  & 0.463  & 4.54 \\
                & K$_{t}$ & 0.404 & 0.426 & 5.45  & 0.480  & 0.458 & 4.58 \\
Bilayer       & $\Gamma$ &    &    &    & 1.039 & 1.321  & 27.91 \\
               &  K$_{l}$ & 0.430 & 0.457 & 6.28  &   &   &   \\
              &  K$_{t}$ &  0.423 & 0.454 & 7.33  &   &   &   \\
Bulk        &  $\Gamma$  &    &    &    & 0.785 & 0.917 & 16.82 \\
              &  $\Sigma_{l}$ & 0.574 & 0.712 & 24.04  &   &   &   \\
              &  $\Sigma_{t}$ & 0.819 & 0.999 & 21.98 &   &   &   \\     [0.1ex] % [1ex] adds vertical space
\hline %inserts single line
\end{tabular}
\label{table2} % is used to refer this table in the text
\end{table}
\end{center}
\end{widetext}

\begin{table*}
\caption{Tight-binding parameters for MoS$_2$ using non-orthogonal model with sp$^{3}$d$^5$ orbitals, nearest-neighbour interactions, and spin-orbit coupling: on-site energies ($E$), spin-orbit splitting ($\lambda$), Slater-Koster energy integrals ($E_1$ for intra-layer and $E_2$ for inter-layer interaction) and overlap integrals ($O_1$ for intra-layer and $O_2$ for inter-layer interaction). The energies are in the unit of eV.
}
\ra{1.3}
 %\centering % used for centering table
\begin{tabular}{lrrrrrrrr}
\hline\hline %inserts double horizontal lines
   & $E_s$  & $E_p$  & $E_d$ & $\lambda_{SO}$  && &&\\
\hline % inserts single horizontal line
S  & 7.6595 & -2.1537 & 8.7689 & 0.2129 && &&\\
Mo & 5.5994 & 6.7128  & 2.6429 & 1.0675 && &&\\
\hline % inserts single horizontal line
           & $E_1$(S,Mo) & $E_1$(S,S)  & $E_2$(S,S)  & $E_1$(Mo,Mo) & $O_1$(S,Mo)  & $O_1$(S,S)  & $O_2$(S,S)  & $O_1$(Mo,Mo) \\
\hline % inserts single horizontal line
$ss\sigma$ & -0.0917   &  0.3093   &  0.3207   &   0.1768   &   0.0294   &   -0.0532   &  -0.1430   &  -0.0575 \\
$sp\sigma$ & 0.6656    & -0.9210   & -0.1302   &   1.0910   &   0.1042   &    0.0240   &   0.0196   &   0.0057 \\
$ps\sigma$ & -1.6515   &           &           &            &   0.1765   &&& \\
$pp\sigma$ & 1.4008    &  0.7132   &  0.7053   &  -0.3842   &  -0.1865   &    0.0478   &  -0.0486   &   0.0296 \\
$pp\pi$    & -0.4812   & -0.1920   & -0.0980   &   0.5203   &   0.0303   &   -0.0104   &   0.0117   &   0.0946 \\
$sd\sigma$ & 0.2177    & -0.2016   &  0.1164   &  -0.5635   &  -0.0480   &    0.0946   &   0.0297   &  -0.1082 \\
$ds\sigma$ & -1.0654   &           &           &            &  -0.1432   &&& \\
$pd\sigma$ & -2.8732   & -0.5204   & -0.0334   &  -0.2316   &   0.0942   &    0.0724   &  -0.0087   &   0.0212 \\
$dp\sigma$ &  2.1898   &           &           &            &   0.2002   &&& \\
$pd\pi$    & 0.7739    & -0.1203   & -0.0370   &   0.0582   &   0.0132   &    0.0772   &  -0.0031   &  -0.0448 \\
$dp\pi$    & -1.9408   &           &           &            &  -0.2435   &&& \\
$dd\sigma$ & -3.1425   &  0.8347   & -0.2300   &   0.3602   &   0.0273   &    0.1849   &   0.0060   &  -0.0216 \\
$dd\pi$    &  2.4975   &  0.7434   &  0.0050   &   0.0432   &   0.1940   &   -0.0429   &  -0.0378   &  -0.0285 \\
$dd\delta$ & -0.3703   & -0.1919   & -0.1104   &   0.1008   &   0.1261   &   -0.0333   &   0.0007   &   0.0432 \\
\hline % inserts single horizontal line
\end{tabular}
\label{table3} % is used to refer this table in the text
\end{table*}

In conclusion, we have developed a generic TB model for accurately calculating the band structures of bulk, monolayer and bilayer MoS$_{2}$. Our TB model is based on the Slater-Koster method. For the optimization of the TB parameters, accurate target band structures are obtained using the screened hybrid DFT method. The accuracy of our TB model is verified by comparing the band gaps for different transitions and the effective masses at different band edges against the \textit{ab initio} band structures. One main feature of our TB model is that with only one set of parameters it can reproduce the band structures of MoS$_{2}$ of different structural configurations: from bulk to 2D structures. Our tight-binding model can be easily extended to other TMDC materials that show electronic characteristics similar to MoS$_{2}$. The TB model reported in this paper is useful for simulations of quantum transport in nanoelectronic devices based on the TMDC materials.

\textbf{Acknowledgements.} This work is supported by the University Grant Council (Contract No. AoE/P-04/08) of the Government of HKSAR (FZ, JW), NSERC  (HG) and IRAP (LL, YZ) of Canada. F.Z. would like to thank Dr. Ji Wei and Mr. Darshana Wickramaratne for useful discussions on the VASP calculations.


\vspace{0.5cm}
\begin{thebibliography}{100}

%1
\bibitem{rHeinz}
K. F. Mak, C. Lee, J. Hone, J. Shan, and T. F. Heinz, Phys. Rev. Lett. \textbf{105}, 136805 (2010).

%2
\bibitem{rKis}
B. Radisavljevic, A. Radenovic, J. Brivio, V. Giacometti, and A. Kis, Nat. Nanotechnol. \textbf{6}, 147 (2011).

%3
\bibitem{rLiu}
H. Liu and P. D. Ye,  IEEE Electron Dev. Lett. \textbf{33}, 546 (2012).

%4
\bibitem{rSayeef}
Y. Yoon, K. Ganapathi, and S. Salahuddin, Nano Lett. \textbf{11}, 3768 (2011).

%5
\bibitem{rGuo}
L. Liu, S. B. Kumar, Y. Ouyang, and J. Guo, IEEE Trans. Electron Dev. \textbf{58}, 3042 (2011).

%6
\bibitem{rKalam}
K. Alam and R. K. Lake, IEEE Trans. Electron Dev. \textbf{59}, 3250 (2012).

%7
\bibitem{rRoger}
R. C. Bowen, G. Klimeck, R. Lake, W. R. Frensley, and T. Moise, J. Appl. Phys. \textbf{81}, 3207 (1997).

%8
\bibitem{rGerhard}
M. Usman, H. Ryu, I. Woo, D. S. Ebert, and G. Klimeck, IEEE Trans. Nanotechnol. \textbf{8}, 330 (2009).

%9
\bibitem{rMathieu}
M. Luisier and G. Klimeck, Phys. Rev. B, \textbf{80}, 155430 (2009).

%10
\bibitem{rIntel}
U. E. Avci, S. Hasan, D. E. Nikonov, R. Rios, K. Kuhn, I. A. Young, Symp. VLSI Tech., pp. 183 (2012).

%11
\bibitem{rHeyd1}
J. Heyd, G. E. Scuseria, and M. Ernzerhof, J. Chem. Phys. \textbf{118}, 8207 (2003); \textbf{124}, 219906 (2006).

%12
\bibitem{rKresse1}
Y.-S. Kim, K. Hummer, and G. Kresse, Phys. Rev. B \textbf{80}, 035203 (2009).

%13
\bibitem{rKresse2}
Y.-S. Kim, M. Marsman, and G. Kresse, Phys. Rev. B \textbf{82}, 205212 (2010).

%14
\bibitem{rHyd2}
J. E. Peralta, J. Heyd, and G. E. Scuseria, Phys. Rev. B \textbf{74}, 073101 (2006).

%15
\bibitem{rLamb}
T. Cheiwchanchamnangij and W. R. L. Lambrecht, Phys. Rev. B \textbf{85}, 205302 (2012).

%16
\bibitem{rEllis}
J. K. Ellis, M. J. Lucero,1 and G. E. Scuseria, Appl. Phys. Lett. \textbf{99}, 261908 (2011).

%17
\bibitem{rWalle}
H. Peelaers and C. G. Van de Walle, Phys. Rev. B \textbf{86}, 241401(R) (2012).

%18
\bibitem{rVasp}
G. Kresse and J. Hafner, Phys. Rev. B \textbf{47}, R558 (1993); G. Kresse and J. Furthmuller, Phys. Rev. B \textbf{54}, 11169 (1996).

%19
\bibitem{rPBE}
J. P. Perdew, K. Burke, and M. Ernzerhof, Phys. Rev. Lett. \textbf{77}, 3865 (1996).

%20
\bibitem{rvdW}
T. Bucko, J. Hafner, S. Lebegue, and J. G. Angya, J. Phys. Chem. A \textbf{114}, 11814 (2010).

%21
\bibitem{rbulk}
A. R. Beal and H. P. Hughes, J. Phys. C \textbf{12}, 881 (1979).

%22
\bibitem{Nanoskif}
Nanoskif is a special tool developed by NanoAcademic Technologies Inc. (www.nanoacademic.com) for the optimization of the Slater-Koster energies and/or overlap integrals used in the tight-binding models by fitting a target band structure of a system, or a set of band structures of several systems simultaneously, employing state-of-the-art algorithms. 

%23
\bibitem{SK}
J. C. Slater and G. F. Koster, Phys. Rev. \textbf{94}, 1498 (1954).

%24
\bibitem{Chadi}
D. J. Chadi, Phys. Rev. B \textbf{16}, 790 (1977).

\end{thebibliography}
\end{document}